\documentclass[onecolumn,nofootinbib]{revtex4}
\usepackage{setspace}
\usepackage{amsmath,amssymb}
\usepackage{graphicx}
\usepackage{dcolumn}
\usepackage{bm,color}
\usepackage{hyperref}
\usepackage{accents}
\usepackage{amssymb,float}
\usepackage{amsmath}
\usepackage{multirow}
\usepackage{tikz}
\usepackage{subcaption}
\usepackage[rightcaption]{sidecap}
\sidecaptionvpos{figure}{t}
\usepackage{enumerate}
\renewcommand{\arraystretch}{3}

\usepackage{hyperref}
\hypersetup{
    colorlinks=true,
    linkcolor=blue,
    filecolor=magenta,      
   citecolor=blue
}
\newcolumntype{P}[1]{>{\centering\arraybackslash}p{#1}}
\newcolumntype{M}[1]{>{\centering\arraybackslash}m{#1}}

\newcommand{\lc}[1]{\accentset{\circ}{#1}}

\begin{document}

\title{Revisiting Stability in New General Relativity}

\author{Sebastian Bahamonde}
\email{sebastian.bahamonde@ipmu.jp}
\affiliation{Kavli Institute for the Physics and Mathematics of the Universe (WPI),\\
The University of Tokyo Institutes for Advanced Study (UTIAS),\\
The University of Tokyo, Kashiwa, Chiba 277-8583, Japan.}
\affiliation{Cosmology, Gravity, and Astroparticle Physics Group, Center for Theoretical Physics of the Universe,
Institute for Basic Science (IBS), Daejeon, 34126, Korea.}

\author{Daniel Blixt}
\email{d.blixt@ssmeridionale.it}
\affiliation{Scuola Superiore Merdionale, Largo S. Marcellino 10, I-80138, Napoli, Italy}

\author{Konstantinos F. Dialektopoulos}
\email{kdialekt@gmail.com}
\affiliation{Department of Mathematics and Computer Science, Transilvania University of Brasov, 500091, Brasov, Romania}

\author{Anamaria Hell}
\email{anamaria.hell@ipmu.jp}
\affiliation{Kavli Institute for the Physics and Mathematics of the Universe (WPI),\\
The University of Tokyo Institutes for Advanced Study (UTIAS),\\
The University of Tokyo, Kashiwa, Chiba 277-8583, Japan.}

\date{\today}

\begin{abstract}
\noindent We study the degrees of freedom in New General Relativity -- flat and metric compatible family of theories --  around the Minkowski background in a gauge invariant manner. First, we confirm the decoupling case, in which the theory reduces to linearized gravity plus a massless KR field. We then show that, unless they vanish, the vector modes of this theory will be unstable. In addition, we find two new branches of the theories, which are instability-free and propagate linearly two tensor modes and in one of the cases also a massless scalar field. This shows that while the generic theory is ill-behaved, there are three possible realizations of instability-free cases, in contradiction to the previous literature, which states that there is only one healthy theory in addition to general relativity.
\end{abstract}

\maketitle

\section{Introduction}\label{sec:intro}

\noindent General Relativity (GR)  is one of the most well-tested theories of nature that has gained numerous confirmations. However, recent observations and theoretical studies suggest that this theory is incomplete. Modified theories of gravity -- theories that recover GR only in certain limits seem to be more promising candidates to underline gravitation. Among them, metric-affine gravity is particularly interesting~\cite{Hehl:1994ue}. In contrast to GR, it allows the possibility of nonvanishing torsion and nonmetricity, utilizing and expanding the geometric approach to gravity. With this, one finds a much wider range of possibilities for new types of particles, while at the same time reopening the possibility of having a renormalizable theory~\cite{Melichev:2023lwj}. However, in this framework, one can easily encounter unstable degrees of freedom (dof)~\cite{BeltranJimenez:2019acz}, i.e. modes that classically lead to linear instabilities, while when quantized cause the violation of unitarity. Nevertheless, there are some potential directions to construct ghost-free metric-affine theories~\cite{Mikura:2023ruz,Mikura:2024mji,Bahamonde:2024sqo}. Whenever we say instabilities herein we mean either ghostly instabilities, meaning that either the temporal derivative of a dof in the perturbed Lagrangian will have the opposite sign, or Ostrogradsky instabilities, meaning that there appear higher-than-first-order derivatives in the Lagrangian. 
\\
\\
\noindent In this paper, we will focus on a particular corner of these theories, with vanishing curvature and non-metricity known as teleparallel gravity~\cite{Aldrovandi:2013wha,Bahamonde:2021gfp}. In particular, we will consider the New General Relativity (NGR), a special metric teleparallel theory that departs from GR only by quadratic and parity-preserving torsion contributions~\cite{Hayashi:1979qx}. Perturbatively, around a Minkowski spacetime, this theory describes an antisymmetric two form, known as the Kalb-Ramond (KR) field, and the usual symmetric, massless spin-2 field. For a long time, this theory was believed to have only a single instability-free case, in which the teleparallel equivalent of GR (TEGR) \cite{BeltranJimenez:2019esp} could be recovered, along with the pseudoscalar of the KR field \cite{Hayashi:1979qx,Ortin:2015hya,Kuhfuss:1986rb,VanNieuwenhuizen:1973fi,BeltranJimenez:2019nns,Blixt:2020ekl}. That theory was labeled the 1-parameter NGR theory. However, recent claims in the literature suggest a much wider set of instability-free possibilities \cite{Golovnev:2023ddv,Golovnev:2023jnc}, for which this theory could have both scalar, vector and tensor perturbations, as well as pseudovector and pseudoscalar modes. A generalisation of NGR with nine arbitrary functions of the D'Alembertian is studied in \cite{Koivisto:2018loq}. In this paper, we revisit these claims in a gauge-invariant manner. On the one hand, our results reveal that by considering gauge-invariant variables in NGR, the conclusions of \cite{Golovnev:2023ddv,Golovnev:2023jnc} that NGR is instability-free in the generic case are invalid. On the other hand, we confirm the statement of \cite{Golovnev:2023ddv,Golovnev:2023jnc} that the analysis of \cite{VanNieuwenhuizen:1973fi} does not necessarily imply the presence of instabilities. In particular, we show that there are indeed two new different instability-free sectors of NGR that have been previously overlooked in the literature by the unfounded statement that the theory has instabilities.
\\ 
\\
\noindent The article is structured as follows: In Sec. \ref{sec:NGR} we introduce NGR. In Sec. \ref{sec:Linear} perturbations of NGR around Minkowski are presented. In Sec.~\ref{sec:StrongCoupling} we discuss the question of strong coupling and the viability of the instability-free branches and in Sec.~\ref{sec:conclusions} we present the conclusions. Throughout the paper, we will use the $\eta_{\mu\nu}=\textrm{diag}(-1,1,1,1)$ signature. The Greek and capital Latin indices correspond to spacetime and tangent space indices, respectively, while small Latin indices indicate spatial components. We will write overcircles $\lc{x}$ on top of a quantity, to distinguish Riemannian quantities.

\section{New General Relativity} 
\label{sec:NGR}

\noindent In this section, we will discuss the basics of New General Relativity by first discussing the necessary geometric quantities. This theory can be formulated by using either the metric $g_{\mu\nu}$ or the tetrad $e^A{}_\mu$. For convenience, we will consider the second formulation. The tetrad represents a set of Lorentz frames that carry a total of 16 components in 4 dimensions, the additional 6 of which, compared to the metric are related to the antisymmetric part of it and are removed by the Lorentz invariance. 
\\
\\
 The tetrad field and its inverse $e_A{}^\mu$ satisfy the following relations:
\begin{equation}
    e^A{}_\mu e_A{}^\nu = \delta _\mu ^\nu \qquad \text{and} \qquad e^A{}_\mu e_B{}^\mu = \delta ^A_B\,,
\end{equation}
while the spacetime metric is then given in terms of the tetrad from
\begin{equation}
    g_{\mu\nu} = \eta _{AB} e^A{}_\mu e^B{}_\nu\,,
\end{equation}
with $\eta _{AB}$ being the Minkowski metric. The torsion tensor can be expressed in terms of the tetrad in the Weitzenb\"ock gauge (zero spin connection) as
\begin{equation}
    T^A{}_{\mu\nu} = \partial _\mu e^A{}_\nu - \partial _\nu e^A{}_\mu \qquad \Rightarrow\qquad  T^\alpha {}_{\mu\nu} = e_A{}^\alpha \left( \partial _\mu e^A{}_\nu - \partial _\nu e^A{}_\mu\right)\,. 
\end{equation}
From torsion, we can construct three independent scalars, while at the same time keeping the higher-derivative or parity-violating terms absent: 
\begin{equation}
    T_{\rm axi} = a_\mu a^\mu\, , \qquad    T_{\rm vec} = v_\mu v^\mu\, , \qquad \text{and}\qquad    T_{\rm ten} = t_{\sigma\mu\nu} t^{\sigma \mu\nu}\,.
\end{equation}
Here, $a_\mu, v_\mu$ and $t_{\sigma \mu\nu}$ are the irreducible decompositions of the torsion tensor under the 4-dimensional pseudo-orthogonal group, that are given by: 
\begin{equation}
    a_\mu = \frac{1}{6}\epsilon _{\mu\nu\rho \sigma}T^{\nu\rho\sigma}\,,\qquad  v_\mu = T^\nu{}_{\nu\mu} \,,\qquad\text{and}\qquad  t_{\sigma \mu\nu} = \frac{1}{2}(T_{\sigma \mu\nu} + T_{\mu\sigma \nu}) + \frac{1}{6}( g_{\nu\sigma} v_\mu + g_{\nu\mu} v_\sigma) - \frac{1}{3}g_{\sigma \mu} v_{\nu}\,.
\end{equation}
The Lagrangian of NGR is a linear combination of the above scalars,
\begin{equation}\label{eq:NGR_Lagr}
    \mathcal{L} _{\rm NGR} = c_{\rm axi} T_{\rm axi} + c_{\rm vec} T_{\rm vec} + c_{\rm ten} T_{\rm ten}\,,
\end{equation}
where the particular choices of $c_{\rm axi} = 3/2, c_{\rm vec} = - 2/3$ and $c_{\rm ten} = 2/3$ lead to the Lagrangian of TEGR. One should note that the above Lagrangian can be also written in an alternative form:
\begin{equation}\label{eq:NGR_Lagr-2}
    \mathcal{L} _{\rm NGR} = c_1 T^{\mu\nu\rho}T_{\mu\nu\rho} + c_2 T^{\mu\nu\rho}T_{\rho\nu\mu} + c_3 T^\mu{}_{\mu\rho}T_\nu{}^{\nu\rho}\,,
\end{equation}
where the new constants $c_i$ are related to the previous ones as
\begin{equation}
c_1 = \frac{1}{2}c_{\text{ten}} - \frac{1}{18}c_{\text{axi}}\,, \qquad
c_2 = \frac{1}{2}c_{\text{ten}} + \frac{1}{9}c_{\text{axi}}\,, \qquad\text{and}\qquad
c_3 = c_{\text{vec}} - \frac{1}{2}c_{\text{ten}}\,.
\end{equation}

\noindent By varying the Lagrangian~\eqref{eq:NGR_Lagr} with respect to the tetrad, one can obtain the field equations of the theory, which in vacuum are given by: 
\begin{align}
    &c_{\rm axi} \left( \frac{1}{2}a^\rho a_{(\rho}g_{\mu\nu)} - \frac{4}{9} \epsilon _{\nu\alpha\beta\gamma} a^\alpha t_\mu {}^{\beta\gamma} - \frac{2}{9} \epsilon _{\mu\nu\rho\sigma} a^\rho v^\sigma - \frac{1}{3}\epsilon _{\mu\nu\rho\sigma} \lc{\nabla} ^\rho a^\sigma\right) \nonumber \\
    &+ c_{\rm vec} \left(\frac{1}{2}v^\rho v_{(\rho}g_{\mu\nu)} + \frac{4}{3} t_{\mu [\rho\nu]}v^\rho + 2 g_{\mu[\nu}\lc{\nabla} ^\rho v_{\rho]} - \frac{1}{2} \epsilon_{\mu\nu\rho \sigma} a^\rho v^\sigma \right) \nonumber \\
    &+ c_{\rm ten} \left( \frac{2}{3}t_{\alpha[\beta\gamma]}t^{\alpha\beta\gamma} g_{\mu\nu} - \frac{4}{3} t_{\mu [\rho\sigma ]}t_\nu{}^{\rho\sigma} + 2 \lc{\nabla} ^\rho t_{\mu[\nu\rho ]} -\frac{2}{3} t_{\nu[\mu\rho ]} v^\rho + \frac{1}{2}\epsilon _{\mu\alpha\beta\gamma} a^\alpha t_\nu {}^{\beta \gamma} \right) = 0\,.
\end{align}
The antisymmetric part of these equations is given by: 
\begin{equation}
    \frac{1}{2}\left(\frac{2}{9}c_{\rm axi} + \frac{1}{2}c_{\rm ten} + c_{\rm vec}\right)\lc{\nabla} _\rho T^\rho{}_{\mu\nu} - \frac{1}{2} \left(c_{\rm ten} - \frac{4}{9} c_{\rm axi}\right) \lc{\nabla} _\rho T_{[\mu\nu]}{}^\rho -\frac{1}{2} \left(c_{\rm vec} + \frac{4}{9}c_{\rm axi}\right) T^{\rho\sigma}{}_{[\mu} T_{\nu]\rho\sigma} = 0\,,
\end{equation}
and could be also obtained if we varied the Lagrangian \eqref{eq:NGR_Lagr} with respect to the spin connection if one is not working in the Weitzenb\"ock gauge. One can easily see that for the following combination of constants 
\begin{equation}
    c_{\rm ten} + c_{\rm vec} = 0\qquad \text{and}\qquad   9 c_{\rm ten} -4 c_{\rm axi}=0\,,
\end{equation}
it identically vanishes and gives GR. 
\\\\
The Hamiltonian analysis of NGR was studied in \cite{Blixt:2018znp,Blixt:2019ene,Tomonari:2024ybs} and a classification according to the presence (or not) of primary constraints, depending on the choice of the coupling parameters was also performed. Similarly, in \cite{Guzman:2020kgh} the primary constraints of NGR were classified in the premetric approach. In this paper, we adopt the same name conventions that we present in Table~\ref{Table:name-convention}.
\begin{table}[ht!]
    \centering
  \renewcommand{\arraystretch}{1.5}
    \begin{tabular}{ c || c c }
        Theory  & Parameter space & \# Primary constraints \\ 
        \hline
        \hline
        Type 1 & Generic & 0  \\ \hline
        Type 2 & $c_{\rm vec}=-c_{\rm ten}$ & 3 \\ \hline
        Type 3 & $c_{\rm ten}=\frac{4}{9}c_{\rm axi}$ & 3 \\ \hline
        Type 4 & $c_{\rm ten}=0$ & 5 \\ \hline
        Type 5 & $c_{\rm vec}=0$ & 1 \\ \hline
        Type 6 (TEGR) & $c_{\rm ten}=\frac{4}{9}c_{\rm axi}\,,\,\, c_{\rm vec} = - c_{\rm ten}$ & 6 \\ \hline
        Type 7 & $c_{\rm ten}=\frac{4}{9}c_{\rm axi}\,,\,\,c_{\rm ten} = 0$ &  8 \\ \hline
        Type 8 & $c_{\rm ten}=\frac{4}{9}c_{\rm axi}\,,\,\, c_{\rm vec} = 0$ & 4 \\ \hline
        Type 9 & $c_{\rm ten}=0\,,\,\,c_{\rm vec} = 0$ & 9 \\ \hline
    \end{tabular}
    \caption{Classification of NGR theories according to their primary constraints.}
    \label{Table:name-convention}
\end{table}

\section{Linear Perturbations around the Minkowski spacetime}\label{sec:Linear}
\noindent In this section, we will calculate perturbations around Minkowski spacetime up to second order for the NGR theory, which is given by the Lagrangian~\eqref{eq:NGR_Lagr}. Since the dynamical variable in our formulation of the theory is the tetrad, we will perturb it as
\begin{equation}
    e^A{}_\mu=\delta^A{}_\mu+\delta e^A{}_\mu\,,\label{per}
\end{equation}
where $\delta e^A{}_\mu$ is the tetrad perturbation. Then, the metric tensor up to the first order is given by
\begin{equation}    g_{\mu\nu}=\eta_{\mu\nu}+2\eta_{AB}\delta^A{}_{(\mu}\delta e^B{}_{\nu)}+\mathcal{O}(\delta e^2)\,.
\end{equation}
Furthermore, let us separate the perturbation into its symmetric and antisymmetric parts according to the following expression
\begin{equation}
    h_{\mu\nu}\equiv 2 \eta_{AB}\delta^A{}_{(\mu} \delta e^B{}_{\nu)}\,,\qquad\text{and}\qquad   B_{\mu\nu}\equiv 2 \eta_{AB}\delta^A{}_{[\mu} \delta e^B{}_{\nu]}\,,\label{decom}
\end{equation}
which provides $g_{\mu\nu}=\eta_{\mu\nu}+h_{\mu\nu}+\mathcal{O}(\delta e^2)$. By substituting the relations ~\eqref{per} and~\eqref{decom} into the Lagrangian~\eqref{eq:NGR_Lagr}, the latter becomes up to second order
\begin{equation}
    \mathcal{L}_{\rm NGR}=\mathcal{L}_{hh}+\mathcal{L}_{Bh}+\mathcal{L}_{BB}\,,
\end{equation}
where
\begin{equation}\label{eq:quadratic-Lagrangian}
    \begin{split}
\mathcal{L}_{hh}&=h^{\mu\nu}\left[\frac{3}{2}c_{\rm ten}\Box h_{\mu\nu}+(c_{\rm vec}-2c_{\rm ten})\partial^{\rho}\partial_{\mu}h_{\nu\rho}-\left(c_{\rm vec}-\frac{1}{2}c_{\rm ten}\right)\left(\partial_{\mu}\partial_{\nu}h+\eta_{\mu\nu}(\partial_{\alpha}\partial_{\beta}h^{\alpha\beta}-\Box h)\right)\right]\,,\\
\mathcal{L}_{Bh}&=2(c_{\rm ten}+c_{\rm vec})h^{\mu\nu}\partial_{\mu}\partial^{\rho}B_{\rho\nu}\,,\\
\mathcal{L}_{BB}&=B^{\mu\nu}\left[\left(\frac{1}{2}c_{\rm ten}-\frac{2}{9}c_{\rm axi}\right)\Box B_{\mu\nu}-\left(\frac{4}{9}c_{\rm axi}+c_{\rm vec}\right)\partial_{\nu}\partial^{\rho}B_{\rho\mu}\right]\,,
    \end{split}
\end{equation}
and $\Box=\nabla^\mu\nabla_\mu=\partial^\mu \partial_\mu$. Note that covariant derivatives are equal to partial derivatives since from now we only consider perturbations around the Minkowski spacetime. The action corresponding to the above Lagrangian density has the following gauge-redundancy
\begin{equation}\label{gaugered}
    h_{\mu\nu}\to\Tilde{h}_{\mu\nu}=h_{\mu\nu}-\partial_{\nu}\xi_{\mu}-\partial_{\mu}\xi_{\nu}\quad \text{and}\quad B_{\mu\nu}\to\Tilde{B}_{\mu\nu}=B_{\mu\nu}+\partial_{\mu}\xi_{\nu}-\partial_{\nu}\xi_{\mu}\,.
\end{equation}
 To study the dof of the theory, we will decompose the two fields according to the group of spatial rotations. In particular, similarly to \cite{Hell:2021wzm, Chamseddine:2012gh}, we will decompose the antisymmetric two-form as:
\begin{equation}
    B_{0i}=C_i+\mu_{,i}\,  \qquad \text{with} \qquad C_{i,i}=0\, , \qquad \text{ and }\qquad B_{ij}=\varepsilon_{ijk}(B_k+\chi_{,k})\,  \qquad \text{where}\qquad  B_{i,i}=0\,.
\end{equation}
Here, the comma denotes a derivative with respect to the spatial component, i.e. $X_{,i}=\partial_i X$. Thus, to describe this field, we have a scalar $\mu$ and a pseudoscalar $\chi$, as well as two transverse vector modes $C_i$, and two additional pseudovector ones $B_i$. 
The symmetric spin-2 field can be decomposed as 
\begin{equation}\label{decompositionCPT}
    \begin{split}
        &h_{00}=2\phi\,,\\
        &h_{0i}=B_{,i}+S_i\,,\qquad {\rm with}\qquad S_{i,i}=0\,,\\
        &h_{ij}=2\psi\delta_{ij}+2E_{,ij}+F_{i,j}+F_{j,i}+h_{ij}^{TT}\,,\qquad{\rm with}\qquad F_{i,i}=0\,,\quad h_{ij,i}^{TT}=0\,,\quad h_{ii}^{TT}=0\,.
    \end{split}
\end{equation}
The 10 components of this field are split into one traceless and transverse tensor $h_{ij}^{TT}$, two traceless vectors $S_i,F_i$ and four scalars $\phi,B,\psi,E$. One can easily find that at the linearized level, the (pseudo)scalar, (pseudo)vector, and tensor perturbations decouple. In the following subsections, we will treat these modes separately, and study the dof of the theory.

\subsection{Tensor perturbations}\label{tensor}
\noindent Let us start our analysis with the simplest type of perturbations, the tensor modes. Their corresponding Lagrangian density is given by
\begin{equation}\label{Ltenm}
    \begin{split}
        \mathcal{L}_T=-\frac{3}{2}c_{\rm ten}\partial^{\alpha}h_{ij}^{TT}\partial_{\alpha}h_{ij}^{TT}\,.
    \end{split}
\end{equation}
In the following, we will assume that the starting action has an overall correct signature in front of it when coupled with external matter. Then, the choice
\begin{equation}
    c_{\rm ten}>0\label{condtensor}
\end{equation}
guarantees the absence of tensor instabilities and contains the graviton sector, rendering NGR an extension of GR. In what follows, we will only consider graviational theories, meaning theories with positive, nonvanishing $c_{\rm ten}$ in order for the massless spin-2 field to be present.

\subsection{(Pseudo)scalar perturbations}

\noindent For the scalar modes, we find the following Lagrangian density: 
\begin{equation}\label{Lscm}
    \begin{split}
        \mathcal{L}_S=&4(c_{\rm vec}+c_{\rm ten})\phi\Delta\phi-4\phi\left[(c_{\rm vec}+c_{\rm ten})\Delta\dot B+2(2c_{\rm vec}-c_{\rm ten})\Delta\psi\right]+36c_{\rm vec}\dot\psi\dot\psi+4(c_{\rm ten}+4c_{\rm vec})\psi\Delta\psi\\
        &+(c_{\rm vec}+c_{\rm ten})\left[\dot B\Delta\dot B+\Delta B\Delta B+4 \Delta\dot E\Delta\dot E+4\Delta E\Delta\dot B\right]+24c_{\rm vec}\dot \psi\Delta\dot E+4(5c_{\rm vec}-c_{\rm ten})\psi\Delta\dot B\\
        &+(c_{\rm vec}+c_{\rm ten})\left(\dot\mu\Delta\dot\mu+\Delta\mu\Delta\mu\right)-2(c_{\rm ten}+c_{\rm vec})\mu\Delta\left[\Delta B-2\dot\phi+\Ddot{B}-2\dot\psi-2\Delta\dot E\right]\\&-\left(c_{\rm ten}-\frac{4}{9}c_{\rm axi}\right)\left(\dot\chi\Delta\dot\chi+\Delta\chi\Delta\chi\right)\,.
    \end{split}
\end{equation}
Here, $\Delta=\partial^i\partial_i=\partial_i\partial_i$ is the 3-dimensional Laplacian operator. Depending on the values of the constants that appear in the above Lagrangian, we have different theories as presented in the Table~\ref{Table:name-convention} and the behaviour of the modes will change. In the following analysis, we will consider these cases separately.

\subsubsection{Type 2: $c_{\rm vec}=-c_{\rm ten}$}\label{scalarType2}

\noindent We start with the simplest case, known as Type 2 or as the 1-parameter NGR theory \cite{Cheng:1988zg,BeltranJimenez:2019nns}. It corresponds to the following choice of constants: $c_{\rm vec}=-c_{\rm ten}$, in which case~\eqref{Lscm} becomes: 
\begin{equation}
     \mathcal{L}_S=12c_{\rm ten}\left(-3\dot\psi\dot\psi-\psi\Delta\psi+2\phi\Delta\psi-2\dot\psi\Delta\dot E-2\psi\Delta\dot B\right)-\left(c_{\rm ten}-\frac{4}{9}c_{\rm axi}\right)\left(\dot\chi\Delta\dot\chi+\Delta\chi\Delta\chi\right)\,.
\end{equation}
We see that the scalar modes are decoupled from the pseudoscalar one. Clearly, $\phi$ is a Lagrange multiplier and by varying the action with respect to it, we find
\begin{equation}
    \Delta\psi=0\,,
\end{equation}
which implies that we can freely set $\psi=0$. As a result, the Lagrangian density describes only the pseudoscalar mode, and is given by
\begin{equation}
     \mathcal{L}_S=-\left(c_{\rm ten}-\frac{4}{9}c_{\rm axi}\right)\left(\dot\chi\Delta\dot\chi+\Delta\chi\Delta\chi\right)\,.
\end{equation}
Following our conventions, the condition
\begin{equation}
    c_{\rm ten}-\frac{4}{9}c_{\rm axi}>0\label{condvectorType2}\,,
\end{equation}
guarantees the absence of scalar instabilities for the Type 2 case. Notably, this pseudoscalar is the degree of freedom of the KR field, an antisymmetric two form. As we will see in the upcoming sections, the vector modes for this case are vanishing. Thus, this case describes the KR theory as well as linearized gravity, that are decoupled at the lowest order in the perturbation theory.\footnote{In the parametrization of Eq.~\eqref{eq:NGR_Lagr-2}, this case corresponds to $2c_1 + c_2 + c_3 = 0$ \cite{BeltranJimenez:2019nns}.} 
\\\\
The following choice of constants $ c_{\rm ten}=(4/9)c_{\rm axi}$,
makes the pseudo-scalar vanish, while keeping the tensor modes present. Thus, one recovers the TEGR with this choice. Curiously, one can also note that the pseudo-scalar is independent of the $c_{\rm ten}$ parameter. If one makes $c_{\rm ten}$ vanish, the KR field will nevertheless remain in the theory, and could provide an interesting behavior \cite{BeltranJimenez:2019nns}, for example, one could discuss it as a dark matter candidate in the context of massive gravity \cite{Blixt:2023qbg}.

\subsubsection{Type 1: Generic case}

\noindent Let us now consider the general case, in which all parameters are arbitrary. From Eq.~\eqref{Lscm} we notice that the scalar $\phi$ is not propagating and by varying the action with respect to it, we find the constraint
\begin{equation}
    2(c_{\rm vec}+c_{\rm ten})\Delta\phi=\left[(c_{\rm vec}+c_{\rm ten})\Delta\left(\dot B+\dot \mu\right)+2(2c_{\rm vec}-c_{\rm ten})\Delta\psi\right] \Rightarrow \phi=\frac{1}{2}\left(\dot B+\dot \mu\right)+\frac{2c_{\rm vec}-c_{\rm ten}}{c_{\rm vec}+c_{\rm ten}}\psi\,.
\end{equation}
Substituting $\phi$ back to \eqref{Lscm}, we find 
\begin{align}\label{Lscm2}
        \mathcal{L}_S=&(c_{\rm vec}+c_{\rm ten})\Delta B\Delta B-2\Delta B\left[6c_{\rm vec}\dot\psi+(c_{\rm vec}+c_{\rm ten})\left(2\Delta\dot E+\Delta\mu\right)\right]+(c_{\rm vec}+c_{\rm ten})\Delta\mu\Delta\mu\nonumber\\
        &+4\Delta\mu\left[3c_{\rm vec}\dot\psi+(c_{\rm vec}+c_{\rm ten})\Delta\dot E\right] +36c_{\rm vec}\dot\psi \dot\psi-4(c_{\rm vec}+c_{\rm ten})\psi_{,i}\psi_{,i}-4\frac{(2c_{\rm vec}-c_{\rm ten})^2}{(c_{\rm vec}+c_{\rm ten})}\psi\Delta\psi+24c_{\rm vec}\dot\psi\Delta\dot E
        \nonumber\\
        &+4(c_{\rm vec}+c_{\rm ten})\Delta\dot E\Delta\dot E-\left(c_{\rm ten}-\frac{4}{9}c_{\rm axi}\right)\left(\dot\chi\Delta\dot\chi+\Delta\chi\Delta\chi\right)\,.
\end{align}
In this Lagrangian density, we notice that neither $B$ nor $\mu$ are propagating. Varying the corresponding action with respect to $B$, we find
\begin{equation}
    (c_{\rm vec}+c_{\rm ten})\Delta^2 B=\Delta\left(6c_{\rm vec}\dot\psi+(c_{\rm vec}+c_{\rm ten})\left(2\Delta\dot E+\Delta\mu\right)\right) \quad\Rightarrow \quad    B=2\dot E+\mu+\frac{6c_{\rm vec}}{c_{\rm vec}+c_{\rm ten}}\frac{1}{\Delta}\dot\psi\,,
\end{equation}
and after substituting this solution into \eqref{Lscm2}, the quadratic Lagrangian for the scalar perturbations becomes
\begin{equation}\label{eq:scalar_Lagr}
    \mathcal{L}_S = -\left(c_{\rm ten} - \frac{4}{9} c_{\rm axi} \right) \left(\dot\chi \Delta \dot\chi + \Delta \chi\Delta\chi \right) + \frac{36 c_{\rm vec} c_{\rm ten}}{c_{\rm vec} + c_{\rm ten}} \left(\dot\psi \dot\psi - \psi_{,i}\psi_{,i}\right)\,.
\end{equation}
Therefore, we end up with two scalars, among which one is the contribution of the KR field, while the other appears due to the coupling with gravity. One should note that both $\chi$ and $\psi$ are gauge invariant with respect to the transformations (\ref{gaugered}). 
\\
\\
We can see that the following conditions
\begin{equation}
    c_{\rm ten}-\frac{4}{9}c_{\rm axi}>0\,,\quad \textrm{and}\quad \frac{c_{\rm vec}c_{\rm ten}}{c_{\rm vec}+c_{\rm ten}}>0\label{noghost1}
\end{equation}
guarantee the absence of instabilities. Previously, we have seen that the tensor modes are well-behaved for  $c_{\rm ten}>0$. By taking this into account, we can see that tensor and scalar modes will be well-behaved if 
\begin{equation}
      -c_{\rm vec}> c_{\rm ten} >0\,,\quad \frac{9}{4}c_{\rm ten}> c_{\rm axi}\qquad {\rm or} \qquad c_{\rm vec}>0\,,\quad c_{\rm ten}>0\,,\quad \frac{9}{4}c_{\rm ten}>c_{\rm axi}\,.
\end{equation}
The above conditions guarantee the absence of instabilities for the tensor and scalar sectors in the most general case, also known as Type 1.   

\subsubsection{Types 3-9}
\label{sec:scalar-type3-9}

\noindent In the previous subsections we discussed Type 1 and 2 theories. Let us see how the scalar perturbations behave in the rest of the NGR theories. As already mentioned, Type 4, 7 and 9 do not propagate a spin-2 field and we do not consider them in this work. In addition, Type 6 is merely TEGR, and propagates a massless spin-2 field. Thus, we should check only Types 3, 5 and 8.
\\\\
In Type 3, $\chi$ vanishes from the Lagrangian \eqref{eq:scalar_Lagr} and $\psi$ propagates in a healthy way for $-c_{\rm vec}> c_{\rm ten} >0$, or $c_{\rm vec}> 0$ and $c_{\rm ten}>0$. In Type 5, only the first term of the Lagrangian \eqref{eq:scalar_Lagr} is non-zero and $\chi$ will not give an instability for $c_{\rm ten}> (4/9)c_{\rm axi}$. Finally, in Type 8, the quadratic scalar Lagrangian vanishes and no scalar dof will propagate.

\subsection{Vector perturbations}

\noindent Let us now study the most complicated dof -- the vector modes. Their Lagrangian density is given by
\begin{equation}\label{Lvecm}
    \begin{split}
        \mathcal{L}_V=&-(c_{\rm ten}+c_{\rm vec})\dot S_i\dot S_i+3c_{\rm ten}S_{i,j}S_{i,j}+3c_{\rm ten}\dot F_{i,j}\dot F_{i,j}-(c_{\rm ten}+c_{\rm vec})\Delta F_i\Delta F_i+2(c_{\rm vec}-2c_{\rm ten})\dot S_i \Delta F_i\\
        &-(c_{\rm ten}+c_{\rm vec})\dot C_i\dot C_i+\left(c_{\rm ten}-\frac{4}{9}c_{\rm axi}\right)C_{i,j}C_{i,j}-2\left(\frac{4}{9}c_{\rm axi}+c_{\rm vec}\right)\varepsilon_{ijk}\dot C_i B_{k,j}\\&+\left(c_{\rm ten}-\frac{4}{9}c_{\rm axi}\right)\dot B_k\dot B_k-(c_{\rm ten}+c_{\rm vec})B_{i,j}B_{i,j}+2(c_{\rm vec}+c_{\rm ten})\left[\dot C_i(\Delta F_i-\dot S_i)+\varepsilon_{ijk}B_{k,j}(\Delta F_i-\dot S_i)\right]\,.
    \end{split}
\end{equation}
 In the following, we will first study the above theory in a generic case (Type 1), after which we will consider special cases that naturally arise in the above framework.

\subsubsection{Type 1: Generic case}
\noindent From Eq.~\eqref{Lvecm}, we see that $S_i, F_i, C_i$ and $B_i$ are gauge dependent under the transformations \eqref{gaugered}. Thus, to study the general case, we will form gauge-invariant quantities. One can easily show that by substituting
\begin{equation}
    C_i=U_i-S_i\qquad \text{and}\qquad B_i=-\varepsilon_{ijk}\partial_j\left(V_k+F_k\right),
\end{equation}
and then by further defining
\begin{equation}
    D_i=S_i-\dot F_i\,,
\end{equation}
the vector Lagrangian (\ref{Lvecm}) becomes
\begin{equation}\label{Lvecgen}
        \mathcal{L}_V=-c_M\dot U_i\dot U_i-c_DU_i\Delta U_i-c_D\dot V_i\Delta \dot V_i-c_M\Delta V_i\Delta V_i-2(c_M-c_D)\dot U_i\Delta V_i -(3c_{\rm ten}+c_D)D_i\Delta D_i+2c_D D_i\Delta(U_i+\dot V_i)\,,
\end{equation}
where, for simplicity, we introduced the constants $c_M$ and $c_D$ as
\begin{equation}
    c_M=c_{\rm vec}+c_{\rm ten}\qquad\text{and}\qquad c_D=c_{\rm ten}-\frac{4}{9}c_{\rm axi}\,.
\end{equation}
Now the vectors $D_i, U_i$ and $V_i$ are gauge-invariant with respect to the transformations \eqref{gaugered}. One can immediately notice that gauge-invariant variables give us an advantage, which is that $D_i$ is not propagating. Then, by varying the Lagrangian with respect to it, we find the constraint
\begin{equation}\label{eqg}
    (3c_{\rm ten}+c_D)\Delta D_i=c_D\Delta(U_i+\dot V_i)\qquad \Rightarrow \qquad D_i=\frac{c_D}{3c_{\rm ten}+c_D}(U_i+\dot V_i)\,,
\end{equation}
where $3c_{\rm ten} + c_D = c_{\rm ten} - (1/9) c_{\rm axi} \neq 0$. Note that the case $9c_{\rm ten} = c_{\rm axi} $  propagates unstable modes\footnote{The case where $9c_{\rm ten} = c_{\rm axi} $ leads to $U_i = - \dot{V}_i$ and the Lagrangian \eqref{eqg} becomes
\begin{equation}\label{LvecgenB}
      \mathcal{L}_V=-c_M\ddot V_i\ddot V_i-2c_D\dot V_i\Delta \dot V_i-c_M\Delta V_i\Delta V_i+2(c_M-c_D)\ddot V _i\Delta V_i =-c_M(\ddot V_i\ddot V_i+2\dot V_i\Delta \dot V_i+\Delta V_i\Delta V_i)\,,
\end{equation}
which clearly propagates unstable modes.} Substituting $D_i$ back to \eqref{Lvecgen}, we find 
\begin{equation}\label{eq:vector Lagr type I}
    \mathcal{L}_V=\frac{1}{2c_\mathrm{ten}+c_D}\left[-a\dot U_i\dot U_i-bU_i\Delta U_i-b\dot V_i\Delta\dot V_i-a\Delta V_i\Delta V_i-2 (a-b)\dot U_i\Delta V_i\right]\,,
\end{equation}
where we have further defined
\begin{equation}
    a=c_M(3c_{\rm ten}+c_D)\qquad\text{and}\qquad b=3c_Dc_{\rm ten}\,.\label{eqab}
\end{equation}
Notice that for $a=b$ the two fields decouple\footnote{This condition corresponds to $$c_{\rm vec}=\frac{c_{\rm ten} (8 c_{\rm axi}+9 c_{\rm ten})}{4 (c_{\rm axi}-9 c_{\rm ten})}\,,\qquad \textrm{or}\qquad c_{\rm axi}=c_{\rm ten}=0\,.
$$ For the first one, the theory contains unstable modes, while for the second one, there is no graviton}. However, the relative sign in front of them is different, suggesting that $U_i$ is an instability.

In order to decouple the modes for $a\neq b$ we have to use the field equations of the vectors.  By varying the Lagrangian with respect to $U_i$ and $V_i$, we find the following equations respectively 
\begin{equation}\label{Uieom}
    a\Ddot{U}_i-b\Delta U_i+(a-b)\Delta\dot V_i=0\,,
\end{equation}
and 
\begin{equation}\label{Vieom}
    b\Ddot{V}_i-a\Delta V_i-(a-b)\dot U_i=0\,.
\end{equation}
In order to decouple the two fields, we can play with their equations of motion. In particular, if we multiply \eqref{Uieom} with $(a-b)$ and act on \eqref{Vieom} with a time derivative and then multiply it by $a$, the sum of the two equations will give
\begin{equation}\label{DeltaUi}
    b(a-b)\Delta U_i=(a-b)^2\Delta\dot V_i-a^2\Delta\dot V_i+ab\partial_0^3V_i\,.
\end{equation}
Now, by acting on \eqref{Vieom} with $\Delta$, and substituting the expression in Eq.~\eqref{DeltaUi}, we find
\begin{equation}\label{eq:vector ghosts}
    \Box^2 V_i=0,\qquad\qquad \text{where}\qquad\qquad \Box^2=\left(\partial_{\mu}\partial^{\mu}\right)^2=\partial_0^4-2\partial_0^2\Delta+\Delta^2\,.
\end{equation}
This indicates that probably there will be two healthy and two unstable vector modes. 

To explore stability, it is convenient to go to the Fourier space:
\begin{equation}
    V_i=\sum_{\sigma=1,2}\int \frac{d^3k}{(2\pi)^\frac{3}{2}}\epsilon^i_{\bf{k}\sigma}v_{\bf{k}}(t)e^{ibf{k}\bf{x}}\qquad \qquad  U_i=\sum_{\sigma=1,2}\int \frac{d^3k}{(2\pi)^\frac{3}{2}}\epsilon^i_{\bf{k}\sigma}u_{\bf{k}}(t)e^{i\bf{k}\bf{x}}.
\end{equation}
Here, $\epsilon^i_{\bf{k}\sigma}$ are the two transverse polarisation vectors, that satisfy
\begin{equation}\label{3polvec}
    \vec{\epsilon}_{\bf{k}\sigma}\vec{\epsilon}_{\bf{k}\sigma'}=\delta_{\sigma\sigma'},\qquad \vec{k}\cdot\vec{\epsilon}_{\bf{k}\sigma}=0 \qquad\text{and}\qquad \sum_{\sigma=1}^2\epsilon^i_{\bf{k}\sigma}\epsilon^j_{\bf{k}\sigma}=\delta^{ij}-\frac{k^ik^j}{|\vec{k}|^2}.
\end{equation}
One easily finds that the momentum space Lagrangian density $L=\int dt d^3k \mathcal{L}_k$, quadratic in vector perturbations, becomes then: 
\begin{eqnarray}\label{Lkvecmod}
    \mathcal{L}_k=\frac{1}{2c_\mathrm{ten}+c_D}\left[-a\dot u_{\bf{k}}\dot u_{-\bf{k}}+bk^2u_{\bf{k}}u_{-\bf{k}}+bk^2\dot v_{\bf{k}}\dot v_{-\bf{k}}-ak^4 v_{\bf{k}} v_{-\bf{k}}+ (a-b)k^2\left(\dot u_{\bf{k}} v_{-\bf{k}}+\dot u_{-\bf{k}} v_{\bf{k}}\right)\right]\
\end{eqnarray}
Let us now explore the stability of the Type I by analysing the Hessian matrix (kinetic) associated to the Lagragian~\eqref{eq:vector Lagr type I}. Then, the kinetic matrix can be written as
\begin{align}
\mathcal{K}_{IJ} = \frac{\partial^2\mathcal{L}_k}{\partial \dot{\Phi}^I \dot{\Phi}^J} 
= \frac{1}{2c_\mathrm{ten} + c_D}
\left(
\scalebox{1.4}{
$\begin{smallmatrix}
-a & 0     & 0     & 0 & 0 & 0 \\
0      &-a & 0     & 0 & 0 & 0 \\
0      & 0      & -a & 0 & 0 & 0 \\
0      & 0      & 0      & k^2 b & 0 & 0 \\
0      & 0      & 0      & 0 &  k^2 b & 0 \\
0      & 0      & 0      & 0 & 0 & k^2 b \\
\end{smallmatrix}$
}
\right) \,,
\end{align}
where $\Phi_I=\{u,v\}$ and whose determinant becomes
\begin{eqnarray}
  \textrm{det} \,\mathcal{K}_{IJ}=-\frac{a^3 b^3 k^6}{(c_D+2 c_\mathrm{ten})^6} \,. 
\end{eqnarray}
The Ostrogradski theorem states that the only way to avoid instabilities in a system with such terms is if the determinant equals zero~\cite{ostrogradsky1848memoire,Delhom:2022vae}. This implies that the only way to avoid instabilities is by setting either \(a = 0\) or \(b = 0\). Using the definitions in equation~\eqref{eqab}, we find that the only way to avoid instabilities is in the following cases:
\begin{itemize}
    \item \(c_M = 0\) or equivalently \(c_{\rm vec} = -c_{\rm ten}\): This corresponds to the Type 2 case, which was already studied in the previous sections.
    \item \(c_{D} = 0\) or equivalently \(c_{\rm ten} = \frac{4}{9}c_{\rm axi}\): This corresponds to the Type 3 case, which was also already studied in the previous sections.
    \item \(3c_{\rm ten} + c_D = 0\) or equivalently \(c_{\rm ten} = \frac{1}{9}c_{\rm axi}\): In this case, the vector sector is stable, but the other sectors (scalar and tensor) cannot be stable. This is evident from the stability conditions~\eqref{condtensor} and~\eqref{noghost1}.
    \item \(c_{\rm ten} = 0\): This case is discarded because it would imply a non-propagating graviton, which we aim to avoid.
\end{itemize}
Therefore, we conclude that the generic Type 1 case will always contain instabilities. An alternative way to see this is by computing the Hamiltonian and verifying whether it is unbounded from below. For this, it is convenient to first rescale fields as
\begin{eqnarray}
    \Tilde{u}_{\bf{k}}=\frac{1}{\sqrt{2c_{\rm ten}+c_M}}u_{\bf{k}}\qquad\qquad  \Tilde{v}_{\bf{k}}=\frac{1}{\sqrt{2c_{\rm ten}+c_M}}v_{\bf{k}},
\end{eqnarray}
after which (\ref{Lkvecmod}) becomes: 
\begin{eqnarray}
    \mathcal{L}_k=\frac{1}{2c_\mathrm{ten}+c_D}\left[-a\dot{\Tilde{u}}_{\bf{k}}\dot{\Tilde{u}}_{-\bf{k}}+bk^2\Tilde{u}_{\bf{k}}\Tilde{u}_{-\bf{k}}+bk^2\dot{\Tilde{v}}_{\bf{k}}\dot{\Tilde{v}}_{-\bf{k}}-ak^4 \Tilde{v}_{\bf{k}} \Tilde{v}_{-\bf{k}}+ (a-b)k^2\left(\dot{\Tilde{u}}_{\bf{k}} \Tilde{v}_{-\bf{k}}+\dot{\Tilde{u}}_{-\bf{k}} \Tilde{v}_{\bf{k}}\right)\right]\
\end{eqnarray}
The conjugated momentas in the momentum space are given by: 
\begin{eqnarray}
    \pi_{U\bf{k}}=-2a\dot{\Tilde{u}}_{\bf{k}}+2(a-b)k^2\Tilde{v}_{\bf{k}}\qquad\qquad \pi_{V\bf{k}}=2bk^2\dot{\Tilde{v}}_{\bf{k}}
\end{eqnarray}
Therefore, the corresponding Hamiltonian density in the momentum space is given by: 
\begin{eqnarray}
    \mathcal{H}_k=-\frac{\left(k^{2} \left(-a +b \right) \Tilde{v}_{\bf{k}}+\frac{ \pi_{U\bf{k}}}{2}\right)^{2}}{a}+a \,k^{4} \Tilde{v}_{\bf{k}}\Tilde{v}_{-\bf{k}}+\frac{\pi_{V\bf{k}}^{2}}{4 b \,k^{2}}-b \,k^{2} \Tilde{u}_{\bf{k}}\Tilde{u}_{-\bf{k}}
\end{eqnarray}

We can notice that due to the relative sign in front of coefficients $a$ and $b$ between the first two and the last two terms respectively, we will always have a Hamiltonian that is unbounded from below.

\subsubsection{Type 2: $c_{\rm vec}=-c_{\rm ten}$}

\noindent Let us now consider the case in which the symmetric and antisymmetric part of the theory decouple. We can notice that the previously given gauge-invariant variables connect the two types of fields. Therefore, in order to study this case, we will rather consider the original variables, given in \eqref{Lvecm}, that now becomes
\begin{equation}\label{eq:Lvector type II}
        \mathcal{L}_V=3c_{\rm ten}\left[ S_{i,j}S_{i,j}+\dot F_{i,j}\dot F_{i,j}-2\dot S_i \Delta F_i\right] +\left(c_{\rm ten}-\frac{4}{9}c_{\rm axi}\right) \left[ C_{i,j}C_{i,j} +2 \varepsilon_{ijk}\dot C_i B_{k,j} +\dot B_k\dot B_k\right]\,.
\end{equation}
The first part of the above Lagrangian density corresponds to the vector part of the linearized gravity, while the second corresponds to the massless KR field. We can clearly see that the above theory has two constraints. The first, associated with $S_i$ is given by
\begin{equation}
    \Delta S_i=\Delta\dot F_i \qquad \Rightarrow \qquad S_i=\dot F_i\,,
\end{equation}
while the second one can be obtained by varying the corresponding action with respect to $C_i$
\begin{equation}
    \Delta C_i=-\varepsilon_{ijk}\dot{B}_{k,j} \qquad \Rightarrow \qquad C_i=\frac{-\varepsilon_{ijk}}{ \Delta}\dot{B}_{k,j}\,.
\end{equation}
Here, $1/\Delta$ corresponds to $|\Vec{k}|^{-2}$ in the Fourier space. Substituting them back to \eqref{eq:Lvector type II}, one finds that all terms cancel, i.e. $\mathcal{L}_V=0$, thus implying that there are no vector modes in this case. Therefore, Type 2 (the decoupling limit), summarized by equations~\eqref{condtensor} and~\eqref{condvectorType2}, corresponds exactly to linearized gravity and the massless KR field.

\subsubsection{Type 3: $c_{\rm ten}=\displaystyle\frac{4}{9}c_{\rm axi}$}
\noindent In addition to the decoupling limit, another interesting case is $c_{\rm ten}=(4/9)c_{\rm axi}$. Under this assumption, from~(\ref{Lvecm}), one can see that the $B_i$ component is not propagating and satisfies the following constraint:
\begin{equation}
    \Delta B_k=-\varepsilon_{ijk}\partial_j\left(\dot C_i-\Delta F_i+\dot S_i\right)\,.
\end{equation}
By solving it and substituting back to (\ref{Lvecm}), we find
\begin{equation}\label{type3vec}
    \mathcal{L}_V=3c_{\rm ten}\left(-\dot F_i\Delta \dot F_i-2\dot S_i\Delta F_i-S_i\Delta S_i\right)\,.
\end{equation}
We can notice that now the $S_i$ component is constrained, in the same manner as in linearized gravity. Thus, substituting $S_i=\dot{F}_i$ back to the above expression, we can see that the overall Lagrangian density for the vector modes vanishes, $\mathcal{L}_V=0$. Notably, in this case, the tensor modes are still present in the theory, as well as the scalar $\psi$, while the pseudo-scalar $\chi$ vanishes. Therefore, we have found a case that propagates a healthy scalar and two healthy tensor modes, that to date, was not carefully considered in the literature apart from recent work \cite{Golovnev:2023ddv,Golovnev:2023jnc}.

\subsubsection{Type 5: $c_{\rm vec} = 0$ and Type 8: $c_{\rm ten} = \displaystyle\frac{4}{9}c_{\rm axi}$ and $c_{\rm vec} = 0$}
\noindent Type 5 is similar to the generic Type 1, except for the presence of an additional symmetry for which the scalar field $\psi$ is non-propagating. Then, the quadratic Lagrangian for the vector perturbations will be \eqref{eq:vector Lagr type I}, but with $c_{\rm vec} = 0.$  This condition though, is not enough to change the behaviour of the equation \eqref{eq:vector ghosts} and thus the theory contains unstable modes.
\\\\
Type 8 though, is like Type 3 with the same additional symmetry implied by $c_{\rm vec} = 0$, which makes the scalar field $\psi$ non-propagating. Thus the Lagrangian \eqref{type3vec} 
will not be affected, i.e. it will be zero, and since there is no scalar dof propagating according to Sec.~\ref{sec:scalar-type3-9}, the theory will only propagate two tensor dof for $c_{\rm ten}>0$; just like TEGR. This is a new healthy theory that has been overlooked as well apart from~\cite{Golovnev:2023ddv,Golovnev:2023jnc}.

\section{Healthy branches and strong coupling}\label{sec:StrongCoupling}
\noindent In the previous sections, we have explored all cases of NGR, which propagate a spin-2 tensor field. We have found that unless vanishing, the vector modes will be unstable. Following the naming convention in Sec.~\ref{sec:NGR}, our results are summarised in Table~\ref{Table:ghostsI}:
\begin{table}[H]
\captionsetup{justification=raggedright, singlelinecheck=false}
\renewcommand{\arraystretch}{1.3}
\centering
\begin{tabular}{ c || c | c | c | c }
    \hline
    \multirow{2}{*}{Theory} & \multirow{2}{*}{Parameter space} & Linear & Non-linear & \multirow{1}{*}{Stability} \\
     & & propagating dof & propagating dof & condition \\
    \hline \hline
    \multirow{2}{*}{ Type 1} &  \multirow{2}{*}{Generic} & 8  &  \multirow{2}{*}{8} &  \multirow{2}{*}{Impossible} \\
    &&$\{h^{TT}_{ij},\chi,\psi,U_i,V_i\}$ & &
  \\   \hline
    \multirow{2}{*}{Type 2} & \multirow{2}{*}{$c_{\rm vec}=-c_{\rm ten}$} & 3 & \multirow{2}{*}{6} & $c_{\rm ten} > \frac{4}{9}c_{\rm axi}$\\ 
    & & $\{h^{TT}_{ij},\chi\}$& & $\text{and } c_{\rm ten}>0$ \\
    \hline
    \multirow{2}{*}{Type 3} & \multirow{2}{*}{$c_{\rm ten}=\frac{4}{9}c_{\rm axi}$} &3  & \multirow{2}{*}{5} & $c_{\rm vec}>0$ and  $c_{\rm ten}>0$ \\ 
    & &$\{h^{TT}_{ij},\psi\}$ & & or $-c_{\rm vec}> c_{\rm ten}>0$ \\
    \hline
    \multirow{2}{*}{Type 5} & \multirow{2}{*}{$c_{\rm vec}=0$} &7 & \multirow{2}{*}{7} & \multirow{2}{*}{Impossible} \\ 
    & & $\{h^{TT}_{ij},\chi,U_i,V_i\}$& &   \\
    \hline
    \multirow{2}{*}{ Type 6 (TEGR)} &  $c_{\rm ten}=\frac{4}{9}c_{\rm axi},\,$ & 2  &   \multirow{2}{*}{ 2} &   \multirow{2}{*}{ $c_{\rm ten}>0$} \\  
    &$ c_{\rm vec} = - c_{\rm ten}$ & \,$\{h^{TT}_{ij}\}$ &&\\ \hline
    \multirow{2}{*}{  Type 8} & $c_{\rm ten}=\frac{4}{9}c_{\rm axi},$ & 2 &   \multirow{2}{*}{ 4 or 6} &  \multirow{2}{*}{ $c_{\rm ten}>0$} \\ & $c_{\rm vec} = 0$  & $\{h^{TT}_{ij}\}$&& \\ \hline
\end{tabular}
\caption{Degrees of freedom for different types of NGR and necessary conditions for the theory being instability-free. For comparison, we have added a column related to the propagating dof of the theory in the nonlinear regime using Hamiltonian analysis found in~\cite{Tomonari:2024ybs}.}
\label{Table:ghostsI}
\end{table}

We have shown that NGR has three cases that propagate tensor modes and are at the same time instability-free. The first among these is well known -- the Type 2 case, in which the coupling between the symmetric and anti-symmetric part of the metric vanishes, and whose Lagrangian density is given by
\begin{equation}
    \begin{split}
        \mathcal{L}_{\rm Type \,2}=-\left(c_{\rm ten}-\frac{4}{9}c_{\rm axi}\right)\left(\dot\chi\Delta\dot\chi+\Delta\chi\Delta\chi\right)-\frac{3}{2}c_{\rm ten}\partial^{\alpha}h_{ij}^{TT}\partial_{\alpha}h_{ij}^{TT}\,.
    \end{split}
\end{equation}
Here, $\chi$ is the dof of the KR field, while the tensor modes match with those of linearized gravity. 
In addition, Type 3 is shown to be healthy at the linear level as well. Its Lagrangian density reads 
\begin{equation}
        \mathcal{L}_{\rm Type \,3}=-\frac{36c_{\rm vec}c_{\rm ten}}{c_{\rm vec}+c_{\rm ten}}\partial_{\alpha}\psi\partial^{\alpha}\psi-\frac{3}{2}c_{\rm ten}\partial^{\alpha}h_{ij}^{TT}\partial_{\alpha}h_{ij}^{TT}\,,
\end{equation}
and for the conditions shown in Table~\ref{Table:ghostsI}, the theory propagates a massless spin-2 and a massless scalar dof, which is not the KR field. Finally, the two remaining stable cases are Type 6 (TEGR) and Type 8, both of which describe only the two healthy tensor modes, and can be summarized in the following way
\begin{equation}
    \mathcal{L}_{\rm TEGR/Type\,8}=-\frac{3}{2}c_{\rm ten}\partial^{\alpha}h_{ij}^{TT}\partial_{\alpha}h_{ij}^{TT}\,.
\end{equation}
The above cases are particularly interesting in the limiting cases of the parameter space. In the limit when $c_{\rm ten}\to(4/9)c_{\rm axi}$, the KR scalar remains in the theory for the Type 2 case.  On the other hand, it is absent in Types 3, 6, and 8, implying a discontinuity in the number of dof at linear order between Type 2 and Types 3, 6, and 8.
The scalar $\psi$ is even more intriguing. In the limits $c_{\rm vec}\to 0$, or $c_{\rm vec}\to -c_{\rm ten} $, it remains in the Type 3 theory, but is absent in Type 2, 6 and 8. Originally, a similar discontinuity in the number of dof was noticed in massive linearized gravity, and massive Yang-Mills theory, with mass added ``by hand'' \cite{vanDam:1970vg, Zakharov:1970cc, Slavnov:1972qb, Fierz:1939ix, Glashow:1961tr}. It yielded predictions that could exclude these theories as possible theories of nature, due to the different numbers of the dof in massive and the corresponding massless theories. 
However, as pointed out in \cite{Vainshtein:1971ip, Vainshtein:1972sx}, and subsequently shown in \cite{Deffayet:2001uk, Gruzinov:2001hp, Hell:2021oea}, once the non-linear-terms become of the same order as the linear ones, the modes that cause the discontinuity become strongly coupled and decouple from the remaining dof, up to small corrections that disappear in the massless limit. This mechanism, known as the ``Vainshtein mechanism'' was further extended to other massive gauge theories, and modified theories of gravity \cite{ Chamseddine:2010ub, Alberte:2010it, Mukohyama:2010xz,  Chamseddine:2012gh, Heisenberg:2014rta, Chamseddine:2018gqh, Hell:2021wzm, deRham:2010ik, deRham:2010kj, Dvali:2006su, Hell:2023mph, Hell:2024xbv, Mortsell:2015exa, Hogas:2021lns,Caravano:2021aum}. 
A similar mechanism might hold in NGR as well, for the pseudo-scalar of the Type 2 theory, as well as the scalar in the Type 3. 
\\ \\
Following \cite{Chamseddine:2012gh, Hell:2023mph}, let us estimate the scale at which these scalars might become strongly coupled. The minimal amplitude of quantum fluctuations for the scalar, pseudoscalar, and tensor modes for scales $k$ corresponding to length scales $L\sim1/k$ that is given by \cite{Mukhanov:2007zz}
\begin{equation}
    \left.\delta\psi\right|_{k\sim\frac{1}{L}}\sim\sqrt{\frac{c_{\rm vec}+c_{\rm ten}}{c_{\rm vec}c_{\rm ten}}}\frac{1}{L}\,,\qquad \left.\delta\chi\right|_{k\sim\frac{1}{L}}\sim\sqrt{\frac{1}{c_{\rm ten}-\frac{4}{9}c_{\rm axi}}}\,,\qquad\text{and}\qquad \left.\delta h_{ij}^{TT}\right|_{k\sim\frac{1}{L}}\sim \frac{1}{\sqrt{c_{\rm ten}}L}\,.
\end{equation}
Let us now consider different limits, noting that the overall action corresponding to the NGR theory should be multiplied by the square of the Planck mass $M_{\rm Pl}$ to obtain the correct units. The perturbation theory will break down once the tetrad perturbations become of the order of unity. This means that the pseudo-scalar $\chi$ appearing in the Type 2 theory will become strongly coupled at the scale
\begin{equation}
    L_{\chi}\sim\frac{1}{M_{\rm Pl}\sqrt{c_{\rm ten}-\frac{4}{9}c_{\rm axi}}}\,.
\end{equation}
Beyond this, it should decouple from the tensor modes. In Type 3, the perturbative series for the scalar $\psi$ will break down once it reaches the scale 
\begin{equation}
    L_{\psi}\sim\frac{1}{M_{\rm Pl}}\sqrt{\frac{c_{\rm vec}+c_{\rm ten}}{c_{\rm vec}c_{\rm ten}}}\,.
\end{equation}
Finally, in the $c_{\rm ten}\to 0$ limit, we will have $\delta\psi\sim\delta h_{ij}^T$, and thus we can see that, both scalar and tensor modes will become strongly coupled at  $L\sim\left(\sqrt{c_T}M_{\rm Pl}\right)^{-1}$. This is similar to the strong coupling appearing in the linearized gravity \cite{Hell:2023mph}. While here we have estimated the strong coupling scale based on the condition that the expansion of the tetrad holds, it will nevertheless be interesting to investigate if these scales will change upon taking into account the non-linear terms. 
\\\\ 
It should be pointed out that while here we have considered only a flat background, gravitational theories admit more complicated, curved ones as well. This brings forth an additional important classification of the dof of gravitational theories with respect to the strong coupling. In particular,  if the strongly coupled dof are present around one background but absent around another or appears only nonlinearly, they are called strongly coupled dof. In order to find the correct predictions for strongly coupled dof, linear simple perturbations are not sufficient. As we have previously discussed, this type of strong coupling usually implies a screening mechanism, which can be used to recover the Newtonian limit at Solar system scales and provide interesting phenomenology at larger scales.Note that Types 1, 2, 3, and 6 have a well-defined Newtonian limit, as it is always possible to express those theories as TEGR plus additional corrections. Conversely, Types 5 and 8 lack a clear Newtonian limit because \( c_{\text{vec}} = 0 \), implying that TEGR cannot be recovered in any manner.
\\\\
Type 2 theory has strongly coupled dof. The linear analysis around Minkowski reveals a massless KR field, in addition to the massless spin-2 field. By going to higher order in perturbation it has been shown that the linearized perturbations contain accidental symmetries that are broken already at cubic interactions \cite{BeltranJimenez:2019nns}. This indicates nonlinear extra degrees of freedom around Minkowski backgrounds, which implies the presence of strongly coupled dof. Furthermore, the Hamiltonian analysis of this theory reveals six dof \cite{Tomonari:2024ybs} (i.e. three in addition to the linear dof revealed by our analysis) and that the primary constraints are not of first-class\footnote{In \cite{Cheng:1988zg} there is also a discussion about bifurcations in the theory.}. The most well-studied teleparallel theory, $f(T)$, suffers from the same pathologies, as can be seen from its Hamiltonian analysis \cite{Blixt:2020ekl,Li:2011rn,Ferraro:2018tpu,Blagojevic:2020dyq} and it has indeed been shown that extra degrees of freedom appear for a nontrivial background \cite{Golovnev:2020nln}. 
\\ 

\noindent Type 3 and Type 8 have also strongly coupled dof \cite{Tomonari:2024ybs}. These theories have previously been disregarded by the false argument that they necessarily will contain instabilities \cite{VanNieuwenhuizen:1973fi,Kuhfuss:1986rb}. The argument was that couplings between the antisymmetric and symmetric perturbations will always generate instabilities, but as we have shown in this paper, this is not the case (at least at the linear level around Minkowski). In the same manner as Type 2, the Type 3 theory also propagates an extra scalar degree of freedom, in addition to the massless spin-2 field. However, the Hamiltonian analysis the three first class primary constraints \cite{Tomonari:2024ybs,Okolow:2013lwa}, resulting in five dof nonlinearly (i.e. two more than the linear analysis reveals). The only other teleparallel theories with only first class constraints that are known are TEGR \cite{Blixt:2020ekl,Maluf:1994ji,Blagojevic:2000qs,Okolow:2013lwa,Ferraro:2016wht} (and other theories equivalent to a known Riemannian theory), and Type 1 NGR \cite{Mitric:2019rop}, which is plagued by instabilities. Thus, to our knowledge, this is the first nontrivial teleparallel theory, which is free from the Ostrodgradsky instability and have first class primary constraints. Although it has two strongly coupled dof, which could be related to a Vainshtein-like screening mechanism. Linear perturbations predict a breeding mode for gravitational waves \cite{Hohmann:2018jso} and deviations from TEGR for the PPN parameters \cite{Hohmann:2023rqn}. However, since they are strongly coupled, this statement has to be revised with the inclusion of nonlinear contributions. Finally, Type 8 is similar to Type 3 with an additional symmetry that manifests linearly by the absence of the scalar field $\psi$. Nonlinearly there is a primary constraint for the trace of the conjugate momenta. For this theory the Hamiltonian analysis revealed a bifurcation \cite{Tomonari:2024ybs} while the counting generically resulted in six dof (4 for the special case) which indicates in both cases a discrepency in number of dof compared to the linear analysis, hence, presence of strongly coupled dof. Additionally, Type 8 has $c_{\rm vec}=0$ which is problematic for cosmology since it is known that $T_{\rm vec}$ is the only scalar in NGR that contributes to flat FLRW spacetimes ($T_{\rm axi}=0 = T_{\rm ten}$ in flat FLRW). This will mean that Type 8 will not be able to correctly reproduce the standard FLRW equations.

\section{Conclusions}
\label{sec:conclusions}

\noindent We have revisited linear perturbations in NGR around a Minkowski background using gauge-invariant variables. Within NGR there are five gravitational theories that go beyond TEGR, namely Type 1-3,5, and 8~\cite{Guzman:2020kgh}. Previously, disputing results have been reported in the literature, where it was claimed in \cite{VanNieuwenhuizen:1973fi} that only Type 2 is instability-free, whereas \cite{Golovnev:2023ddv,Golovnev:2023jnc} claim that all of these theories can be instability-free. Our analysis is summarised in Table~\ref{Table:ghostsI} and is in disagreement with both claims. As suggested in \cite{Golovnev:2023ddv,Golovnev:2023jnc}, the conclusions of \cite{VanNieuwenhuizen:1973fi} are indeed too strict since Type 3 and 8 are manifestly instability-free as well. However, we find that it is impossible to avoid instabilities in the generic (Type 1) case and Type 5. Recently, a new add-on for Mathematica has appeared \cite{Barker:2024juc} dubbed \textit{PSALTer} (Particle Spectrum for Any Tensor Lagrangian), which provides the user with  the spin-projection operators, saturated propagator, bare masses, residues of massive and massless poles and overall unitarity conditions in terms of the coupling coefficients, once it gets the quadratic Lagrangian of any field theory around Minkowski. We also used this software with the Lagrangian \eqref{eq:quadratic-Lagrangian} and verified the above results for all types in Table~\ref{Table:name-convention}.
\\\\
Type 1 inevitably propagates a instabilities in the vector sector, while Type 2 and Type 3 only propagate a massless KR\footnote{A massless KR field only propagate the scalar helicity mode since the KR field is a pseudo-vector.} and massless scalar field, respectively, in addition to the graviton. In other words, at the linearized level these theories propagate three dof. On the other hand, in Type 8 only the graviton is propagating. Among all of these theories, Type 2 has been considered more in the literature \cite{Hayashi:1979qx,Hohmann:2018jso,Hohmann:2023rqn,Asukula:2023akj,Golovnev:2023ddv,Cheng:1988zg,BeltranJimenez:2019nns,Blagojevic:2000qs,Blixt:2018znp,Blixt:2019ene,Okolow:2013lwa}. Nevertheless, an important point that will be investigated in the future work is the strong coupling of the dof for Type 2,3, and 8. Although, Type 8 might have some other problems such as not being able to reproduce cosmologically viable models.
\\\\
Due to the statement of \cite{VanNieuwenhuizen:1973fi} Type 3 has not gained much interest in the literature. It has also been disregarded since the PPN parameters seem to be different from those predicted by GR \cite{Hayashi:1979qx,Hohmann:2023rqn}. However, we argue that the theory is worth revisiting because i) it is instability-free at the linearized level and ii) the methodology of \cite{Hayashi:1979qx,Hohmann:2023rqn} is most likely not correct. The second statement follows from the strong coupling which is claimed to exist in this theory \cite{Golovnev:2023jnc}. This claim should be verified in the future by adopting gauge-invariant variables also around cosmological backgrounds. Independently we have argued in Sec. \ref{sec:StrongCoupling} that results from the Hamiltonian analysis \cite{Tomonari:2024ybs,Okolow:2013lwa} shows that, the full nonlinear theory have two additional dof which are strongly coupled but perhaps easier to understand since the constraints are first class implying a symmetry that we can understand better than in other teleparallel theories\footnote{Interestingly, this indicate that the two first, and perhaps also the third, guiding principles suggested in \cite{Bajardi:2023gkd} applies to NGR of Type 3.}. The presence of strongly coupled dof suggests that there may exist a screening mechanism, since nonlinear terms become dominant and need to be incorporated at certain scales, which could be used to recover the Newtonian limit at Solar system scales. 
\\\\
In the future, it would be interesting to investigate the nature of the nonlinear dof for Type 3 NGR. This can be done by i) going to higher orders in perturbation theory around Minkowski, ii) investigate linear perturbations around other backgrounds, and iii) completing the Hamiltonian analysis by classifying all constraints into first and second classes, which has now been done in the recent article \cite{Tomonari:2024ybs}. Related to this, specific backgrounds should be investigated in the context of revealing how, in detail, the screening mechanism (if it exists) works. NGR of Type 3 is, to our knowledge, the first teleparallel extension to GR with an indication of well-behaved dof. However, there is a zoo of teleparallel theories for which this aspect has not been explored at all.  
\\\\
Moreover, we should note that the kinetic term of $\psi$ is particularly curious. It will be especially interesting to study this case in the presence of external sources, where the theory might yield different observable predictions of NGR in comparison to GR. In addition, recently, a possibility of removing the unstable dof was applied to conformal gravity, in order to remove the unstable tensor modes \cite{Hell:2023rbf}. By applying two sets of boundary conditions for the de Sitter and the Minkowski space-time, it was shown that conformal gravity can reduce to Einstein gravity (with the cosmological constant for the first case). It would be interesting to see if the boundary conditions could be applied to this theory as well, in order to remove the vector instabilities, and study to which theory will this theory go to.

\section*{Acknowledgements}

S.B. is supported by “Agencia Nacional de Investigación y Desarrollo” (ANID), Grant “Becas Chile postdoctorado al extranjero” No. 74220006. K.F.D. was supported by the PNRR-III-C9-2022–I9 call, with project number 760016/27.01.2023. The work of A. H. was supported by World Premier International Research Center Initiative (WPI), MEXT, Japan. This paper is based upon work from COST Action CA21136 {\it Addressing observational tensions in cosmology with systematics and fundamental physics} (CosmoVerse) supported by COST (European Cooperation in Science and Technology).

\bibliographystyle{utphys}
\bibliography{references}

\end{document}